\documentclass{mem}
\usepackage{natbib}\usepackage{txfonts}\usepackage{balance}
\usepackage{graphicx}
\usepackage[a4paper,breaklinks,dvipdfm]{hyperref}
\idline{82}{781}
\begin{document}
\def\teff{$T\rm_{eff }$}
\def\kms{$\mathrm {km s}^{-1}$}

\title{Supernova Propagation in the Circumstellar and Interstellar medium}


\author{
Vikram V. \,Dwarkadas 
}

\offprints{V. V. Dwarkadas}

\institute{
Department of Astronomy and Astrophysics,
Univ of Chicago
5640 S Ellis Ave
Chicago, IL 60637
\email{vikram@oddjob.uchicago.edu}
}

\authorrunning{Dwarkadas}

\titlerunning{SNe in CSM and ISM}

\abstract{We describe the propagation of supernova shocks within the
  surrounding medium, which may be due to mass-loss from the
  progenitor star. The structure and density profile of the ejected
  material and surrounding medium are considered. Shock wave
  interaction with clouds and wind-bubbles, and issues relevant to
  cosmic rays are briefly discussed.

} 

\maketitle{}

\section{Introduction}The propagation of Supernova Remnants (SNRs) 
in the ambient medium has been studied for over 4 decades. Early work
was reviewed by \citet{woltjer72}. Since then there have been several
excellent reviews \citep{chevalier77, om88, chevalier1994, bs95}. The
high energy emission from SNRs was reviewed by \citet{reynolds08}. In
this short presentation, our aim is to review the basic ideas of the
propagation of SN shock waves into the ambient medium, and point out
issues relevant to cosmic-ray acceleration. For further details we
refer the reader to the excellent publications cited above.

\section{Ejecta-Dominated Stage}
The explosion of massive stars ($\ga 8 M_{\odot}$), or the
thermonuclear deflagration and detonation of white dwarfs, leads to
the production of highly supersonic shocks waves that propagate into
the ambient medium. In order to understand the evolution of these
shock waves, one must know the density structure of the ejecta and the
surrounding medium. The density structure of the ejecta in
core-collapse SNe can be approximated by a decreasing power law
\citep{cs89, matzner99} with an index roughly between 9 and 11. Below
a certain velocity the density is assumed roughly constant. The
ambient medium in core-collapse SNe is much more complicated. Close to
the star, the density can be approximated by a wind, whose density
varies as r$^{-2}$ if the wind parameters are constant (although there
is no reason for them to be constant, see \citet{dg11}). The expansion
of power-law ejecta decreasing as r$^{-n}$ into a power-law
circumstellar medium whose density decreases as r$^{-s}$ can be
described by a self-similar solution \citep{chevalier1982a, tm99},
where the radius goes as t$^{(n-3)/(n-s)} \sim t^m$, where $m < 1$ is
called the expansion parameter.  For representative values $n=9$ and
$s=2$ indicative of a wind, $m=0.86$. Thus, the presence of a medium,
however tenuous, restricts free expansion of the shock wave, whose
velocity decreases as $t^{m-1}$. Further from the star, the density
depends on the mass-loss from, and evolution of, the progenitor star,
and may form a wind-blown bubble \citep[][\S
  \ref{sec:windbub}]{Weaver1977}.

Type Ia SNe are assumed to arise from low mass stars, which don't
suffer from considerable mass-loss and don't modify their ambient
medium significantly \citep{badenesetal07}.  A first assumption is to
assume an unmodified constant density medium around Ia's. The ejecta
profile is more complicated. By comparing density profiles obtained by
hydrodynamical simulations of Type Ia explosions, \citet[][hereafter
  DC98]{dc98} showed that an exponential profile is a better
approximation for Type Ia ejecta profiles. The introduction of an
exponential introduces one more variable, and therefore the solution
is no longer self-similar, although it is still scalable, and can be
expressed in terms of a single family of parameters.

\begin{figure}[]
\resizebox{\hsize}{!}{\includegraphics[clip=true]{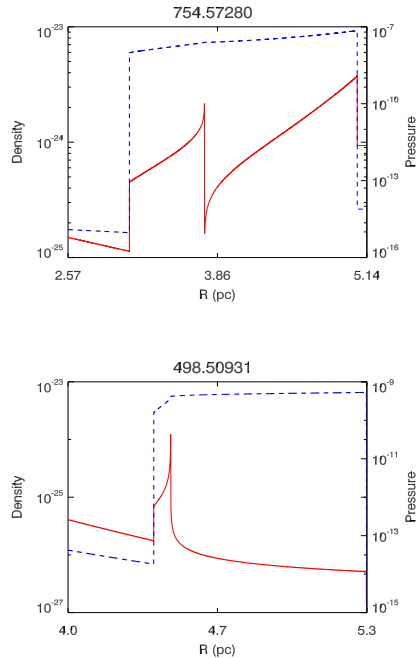}}
\caption{ \footnotesize The density (red) and pressure (purple)
  profiles of the shocked interaction region between the forward and
  reverse shocks, for [Top] a Type Ia SN with an exponential ejecta
  density profile interacting with a constant density medium and
  [Bottom] A core-collapse SN with a power-law density profile
  interacting with a wind with constant parameters. Time in years is
  given at the top. Note the variation in the profile between the two
  cases. An expanding grid is used to allow most of the computational
  domain to be occupied by the shocked region. }
\label{fig:profiles}
\end{figure}

The expansion of the SN ejecta into the ambient medium drives a shock
into the medium. The material swept up by the shock is slower moving,
and the ejecta need to be decelerated before they interact with this
material. This leads to the formation of a reverse shock, that expands
back into the ejecta in a Lagrangian sense. The two are separated by a
contact discontinuity, which divides the region of shocked ejecta from
the region of shocked ambient medium (Fig.~\ref{fig:profiles}). The
shocks will accelerate particles to high energies due to a process
that is generally thought to be some variant of Diffusive Shock
Acceleration.

The self-similar, or scalable solutions, provide the ratio of radii of
the reverse shock, contact discontinuity (CD) and forward shock. These
ratios are fixed for purely non-radiative hydrodynamical evolution,
but may be modified if significant energy ($> 10\%$) is expended in
accelerating particles to relativistic speeds
\citep{caprioli11}. Large modification of these ratios is often taken
as evidence of cosmic-ray acceleration, which would tend to narrow the
distance between the CD and forward shock for instance
\citep{warrenetal05, kosenkoetal11}. However this presumes that the
ejecta profiles, and especially the smooth, homogeneous ambient medium
profiles, are exact representations. For Type Ia's, DC98 showed that
while the exponential is a good and convenient approximation to the
ejecta profiles, deviations from this are important, especially for
low-energy explosions. One would also expect inhomogeneities in the
ambient medium over large scales, especially for core-collapse SNe,
whose effect may remain imprinted in the profile over several doubling
times, and thus modify the ratio. Other factors, including the growth
of R-T instabilities (Orlando, these proceedings), could alter the
distance between CD and forward shock \citep{kosenkoetal11}.

Since the reverse shock is generally expanding into a higher density
than the forward shock, it's velocity is lower. If particles are
accelerated by the reverse shock, as suggested for Cas A \citep{hv08},
they would be expected to have a lower maximum energy than those at
the forward shock. If core-collapse SNe expand in winds whose density
decreases as $r^{-2}$, whereas Type Ia SNe expand into a constant
density medium, then the ratio of the density ahead of the reverse
shock to that ahead of the forward shock falls more rapidly in Type
Ia's then in core-collapse SNe, suggesting that the reverse shock's
effects are felt longer in core-collapse SNe than in Ia's.

During the early stages of the evolution, both the forward and reverse
shocks expand outwards in radius. The decelerating contact
discontinuity is Rayleigh-Taylor (R-T) unstable, and R-T fingers will
mix shocked ejecta material with the shocked ambient medium
\citep{cbe92, dwarkadas00}. At a later time, depending on the ejecta
profile \citep{dc98}, the ejecta density becomes low enough that the
reverse shock will ``turn over'' and actually begin to move inwards in
radius, towards the explosion center. It is only after the reverse
shock reaches the center that the SNR can be thought of as entering
the Sedov-Taylor phase.

\section{Sedov-Taylor Stage} In this adiabatic stage the remnant expansion
depends mainly on two parameters, the energy of the explosion $E$ and
the ambient medium density $\rho$. The shock front travels as $R_{sh}
\propto (E/\rho)^{1/5}\,t^{2/5}$ when expanding into a constant density
medium, or $R_{sh} \propto t^{2/3}$ in the less likely case of the
stage being reached while the SN is expanding in a wind medium with
$\rho \propto r^{-2}$.  This expression was first derived for a point
explosion in a uniform medium \citep{taylor1950}.

In order for the remnant to reach the Sedov stage, several things must
happen (1) The reverse shock must reach the center and effectively
cease to exist, since the Sedov-Taylor solution is a single-shock
solution, as opposed to the double-shocked solutions earlier
(Fig.~\ref{fig:profiles}). In multi-dimensional simulations, the
reverse shock generally does not remain spherical as it moves back,
and eventually tends to dissipate. (2) The expansion parameter $m$
discussed earlier must decrease from its larger value till it reaches
the Sedov value of 0.4 (0.67) in a constant density medium (wind) (3)
The density profiles must change from those of the ejecta-dominated
stage (Fig.~\ref{fig:profiles}) to those of the Sedov stage.  Although
traditionally the Sedov stage is assumed to be reached when the
swept-up mass is just larger than the ejecta mass, this does not hold
in practice, as first shown almost 4 decades ago \citep{gull73}. DC98
showed that the swept-up mass must be about 20-30 times the ejecta
mass (depending on the ejecta profile) when the reverse shock reaches
the center.

Thus the idealistic representation of the Sedov stage is reached later
in time, and SN radius, than when the swept-up mass equals the ejecta
mass. The simple reason is that a SNR is not a point explosion, and
the Sedov stage arises from the ejecta-dominated stage, when the
remnant is already light-years across in radius. Major galactic
remnants, such as Tycho, Cas A, Kepler, and even the 1000-year old SN
1006 are currently in the ejecta-dominated stage, with a
double-shocked structure.

Some calculations suggest that the transformation of SNR energy to
cosmic-rays is most efficient at the beginning of the Sedov stage
\citep{bv97, pz05,pzs10}. This has prompted many investigations into
the acceleration of particles in the Sedov stage \citep{kang06,
  kang10, castroetal11}. The transition into the Sedov stage however
may occupy a significant amount of the lifetime of this stage, and
must be taken into account in computations of the accelerated particle
spectrum.  In some cases, the Sedov stage may be shortened, or even
completely eliminated (\S \ref{sec:windbub}).

\section{Radiative Stage:}  When the shock slows down sufficiently that 
the cooling time of the material behind the shock becomes smaller than
the flow time, the SNR enters the radiative phase. This phase is not
generally of much interest in cosmic-ray physics because the slow
shock is inefficient at accelerating particles to relativistic
energies \citep{bv97}. The evolution can be approximated by a
momentum-driven flow, where $R_{sh} \propto t^{1/4}$. \citet{om88}
prefer a pressure-driven snowplow model ($R_{sh} \propto t^{2/7}$),
while \citet{cmb88} suggest an offset solution $R_{sh} \propto
t^{0.3}$. In any case the SNR has decelerated considerably from the
initial stages. Eventually, it will merge with the ISM, finally mixing
all the constituents of the explosion back into the ISM.

A radiative shock though can form at any time depending on the density
of the ambient medium, if the cooling time $t_{cool} \propto kT_s/(n
\Lambda)$ becomes smaller than the flow time. Here $T_s \propto v_s^2
$ is the postshock temperature, $n$ the density of the ambient medium,
and $\Lambda$ the cooling function. This can happen for young SNe
expanding into a high density medium, as has been postulated for SN
1993J \citep{nymarketal09, chandraetal09}. SNRs impacting with clouds
or driven into clumps may also form radiative shocks. For some SNe,
modelling of the X-ray and optical emission indicates a radiative
shock impacting clumps while an adiabatic shock expands into the
interclump medium \citep{chugai93, chugai94, cdd95}.

\section{Some special Case Studies}
\subsection{SNR-Molecular Cloud Interaction} 
\label{sec:shkcl} 
\begin{figure*}[t!]
\resizebox{\hsize}{!}{\includegraphics[clip=true]{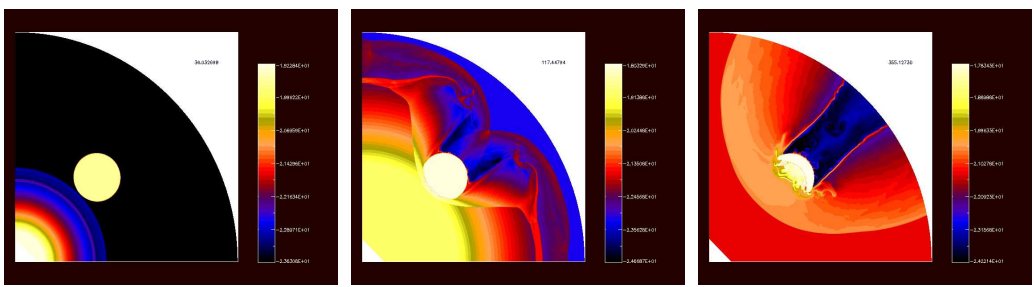}}
\caption{\footnotesize Snapshots from a simulation of the interaction
  of a SNR with a dense cloud, carried out using the VH-1 code in
  spherical symmetry. (a) A SNR with a spherical double-shocked
  structure approaches the cloud. Time in years is at top of
  picture. (b) The impact drives a shock into the cloud, and a
  reflected shock back into the inter-cloud medium. A complicated
  hydrodynamic structure is seen. (c) The rest of the intercloud shock
  has swept past, while the reflected shock can be clearly
  seen. Meanwhile, the shock crossing the cloud results in the
  formation of instabilities that will ultimately destroy the cloud.}
\label{fig:shkcl}
\end{figure*}

This is an interesting case for cosmic ray physicists, because many
SNRs discovered in $\gamma$-rays are found to be interacting with
molecular clouds, suggesting a hadronic origin for the emission. There
are various possibilities, depending on whether the SNR shock is
actually interacting with the cloud or not. (1) Shock-cloud
interaction has not occurred. The escaped particles from the SNR have
interacted with the dense cloud material, leading to broadband
non-thermal emission \citep{gac09}. The hydrodynamical evolution of
the SNR is not yet affected by the presence of the cloud. (2) The SN
shock has impacted the cloud. The impact of SNR shocks with clouds was
discussed by \citet{kmc94}, and resulting $\gamma$-ray emission by
\citet{uchiyamaetal2010}. Fig \ref{fig:shkcl} shows results from a
simulation of a SNR interacting with a cloud of density $\rho_{cl}$.
The impact drives a shock into the cloud, and a reflected shock back
into the intercloud medium with density $\rho_{ic}$. The velocity of
the cloud shock $v_{cl} \propto v_{ic} \,\sqrt(\rho_{ic}/\rho_{cl})$
can be much smaller than the inter-cloud shock velocity. In case of a
dense cloud, the cloud shock can become radiative if it cools faster
than the flow time through the cloud.  The impact of the shock on the
cloud, and the shear flow on the front and sides, leads to the
formation of Richtmeyer-Meshkov and Kelvin-Helmholtz instabilities,
which will eventually lead to the destruction of the cloud. On a
smaller scale, these types of interactions are seen in SNRs
interacting with a clumpy medium.

\subsection{SNRs in Wind-blown Bubbles} 
\label{sec:windbub} 
\begin{figure}[]
\resizebox{\hsize}{!}{\includegraphics[clip=true]{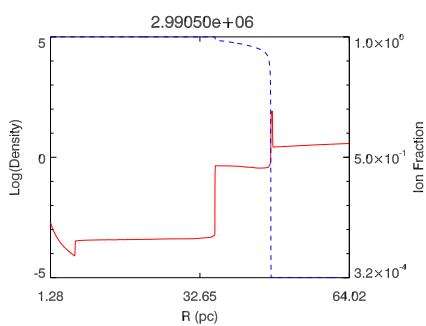}}
\caption{ \footnotesize The basic structure of a wind-blown bubble
  around a main-sequence 40 M$_{\odot}$ star, carried out using the
  ionization-gasdynamics code AVATAR \citep{dr11}. Density in red
  (scale on left), ionization fraction in blue (scale on
  right). Radius in parsecs on X-axis. From left, we see the freely
  expanding wind, the wind termination shock, a shocked wind region,
  an ionized HII region, and the dense shell. The structure upto the
  HII region is fully ionized.  The ionization front is trapped within
  the shell, and the ionization fraction can be seen to drop very
  sharply within the shell.  }
\label{fig:windbub}
\end{figure}

Many massive  stars will form large, and some 
not-so-large, wind bubbles around the star. Wind bubbles are regularly
seen around main-sequence and Wolf-Rayet (W-R) stars
\citep{chu08}. The basic structure (Fig \ref{fig:windbub}) comprises
of an inner and outer shock separated by a contact discontinuity
(CD). The shocked wind between the inner shock and CD usually has a
low density and high pressure, and thus high temperature. The outer
shock is usually radiative, and the material between the outer shock
and CD forms a thin, dense shell. Ionization from the star may result
in the formation of an ionized region around the wind-blown shell.

The post-main-sequence phases occupy a small fraction of the total
stellar lifetime, but may result in substantial mass-loss and
modification of the ambient medium \citep{vanmarleetal05,fhy06,
  Dwarkadas2007c, ta11}.  Stellar evolution theory suggests that
single stars $<$ 30 $M_{\odot}$ will end their lives as RSGs to form
Type IIP SNe. Stars with higher initial mass will become W-R stars,
with perhaps an intermediate Luminous Blue Variable phase, and explode
as Type Ib/c SNe. RSGs have slow dense winds. W-R stars have fast
winds with a lower mass-loss rate (Puls, these proceedings), resulting
in a lower density in the ambient medium into which the SN will
expand. A binary companion can alter the stellar evolution and
structure of the ambient medium, as postulated for SN 1987A
\citep{morris2007}.

The evolution of SN shock waves within these bubbles has been studied
by \citet{Chevalier1989, tenorioetal1990, tenorioetal1991,
  Dwarkadas2005, Dwarkadas2007a, Dwarkadas2007c, ddb10}. The collision
of a SN shock with a shock or CD results in a transmitted shock
expanding outwards, and a reflected shock expanding back into the
ejecta. Thus expansion within a bubble results in several shocks
bouncing around in the remnant cavity. Many of these shocks can
accelerate particles to high energies, at least for short periods of
time. Accelerated particles that escape from the SNR and interact with
this shell may give rise to $\gamma$-ray emission \citep{eb11}. When
the SN shock collides with the dense shell, it can result in
significant deceleration of the shock, and an increase in the optical
and X-ray emission, as seen in SN 1987A \citep{Park2006} and SN 1996cr
\citep{baueretal2008}. The shock will traverse the shell with a low
velocity (see \ref{sec:shkcl}) until it emerges. In extreme cases, the
shell density may be so high that the shock will be trapped in the
shell, and the kinetic energy may be radiated away. The SN will then
go directly to the radiative phase, avoiding the Sedov phase.

\begin{acknowledgements}
I am very grateful to the organizers, especially A.~Marcowith, for
organizing a splendid conference, for inviting me to present this
review, and for their gracious hospitality. It is a pleasure to
acknowledge the many people from whom I have learned about SNe and
SNRs, especially R. Chevalier. My research is supported by grants from
Chandra, and Fermi grant NNX10AO44G.
\end{acknowledgements}

\bibliographystyle{aa}
\bibliography{paper}

\end{document}